# Frequency-agile dual-comb spectroscopy

Guy Millot [1,*], Stéphane Pitois [1], Ming Yan [2,3], Tatevik Hovannysyan [1], Abdelkrim Bendahmane [1], Theodor W. Hänsch [2,3], Nathalie Picqué [2,3,4,†]

1. Laboratoire Interdisciplinaire Carnot de Bourgogne, CNRS Univ. Bourgogne Franche-Comté, 9 avenue Alain Savary, 21078 Dijon, France
2. Ludwig-Maximilians-Universität München, Fakultät für Physik, Schellingstr. 4/III, 80799 Munich, Germany
3. Max-Planck-Institut für Quantenoptik, Hans-Kopfermannstr. 1, 85748 Garching, Germany
4. Institut des Sciences Moléculaires d'Orsay, CNRS, Bâtiment 350, 91405 Orsay, France
* guy.millot@u-bourgogne.fr
† nathalie.picque@mpq.mpg.de

***Abstract :*** *We propose a new approach to near-infrared molecular spectroscopy, harnessing advanced concepts of optical telecommunications and supercontinuum photonics. We generate, without mode-locked lasers, two frequency combs of slightly different repetition frequencies and moderate, but rapidly tunable, spectral span. The output of a frequency-agile continuous wave laser is split and sent into two electro-optic intensity modulators. Flat-top low-noise frequency combs are produced by wave-breaking in a nonlinear optical fiber of normal dispersion. With a dual-comb spectrometer, we record Doppler-limited spectra spanning 60 GHz within 13 µs and 80-kHz refresh rate, at a tuning speed of 10 nm.s$^{-1}$. The sensitivity for weak absorption is enhanced by a long gas-filled hollow-core fiber.*

Sensitive and nonintrusive gas detection finds an increasing number of applications, from biomedical diagnostics to atmospheric sensing. This motivates the exploration of new approaches to optical molecular spectroscopy. Frequency agile laser techniques achieve very fast spectral measurements [1-3]. Concurrently, laser frequency combs [4] open up novel opportunities for precise multiplex [5-13] or multichannel [14-16] spectroscopy with resolved individual comb lines. Each scheme, however, calls for trade-offs: serial time-encoded spectroscopy [1] has a limited precision, dual-comb spectroscopy [5-12] a reduced dynamic range etc. In this letter, we take a different approach, which overcomes some of the drawbacks of frequency-agile and multiplex spectroscopies, and we devise a new scheme involving two mutually coherent frequency-agile frequency combs.

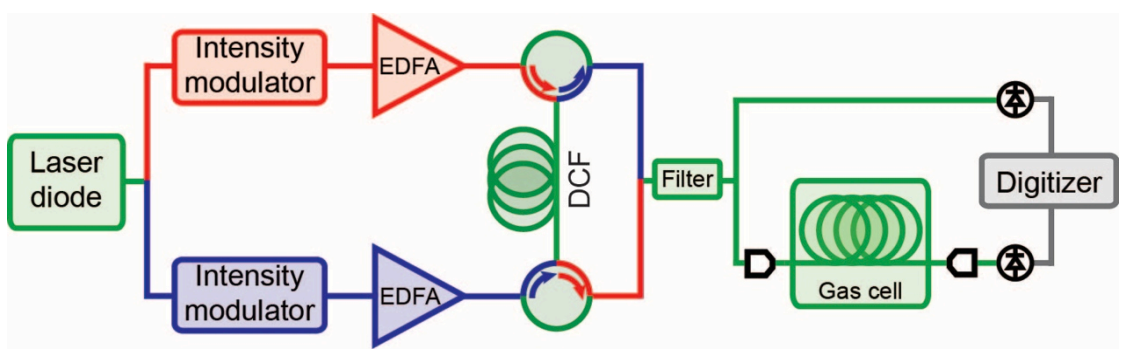

**Figure 1.** Experimental set-up: The continuous-wave laser is intensity-modulated by two electro-optics modulators driven by radio-frequency generators set at slightly different frequencies. Two trains of 50-ps pulses at repetition frequencies of about 300 MHz are spectrally-broadened by wave-breaking in a normal dispersion nonlinear optical fiber, in which they counter-propagate. The gas under study may be contained in a hollow-core photonic crystal fiber, which enhances the absorption path length. The comb beams beat onto a InGaAs fast photodiode. The electric signal is low-pass filtered. The time domain interference signal is then amplified and digitized with a data acquisition board.

Figure 1 sketches the experimental set-up developed in the telecommunication C and L bands (1530-1625 nm, 184-196 THz). A frequency-agile laser diode seeds in parallel two intensity modulators, driven at slightly different repetition frequencies $f$ = 300 MHz and $f+\delta f$ = 300.1 MHz. Their spectrum (Fig.2a) has a full-width at half-maximum (FWHM) of 0.13 nm (16 GHz). The FWHM of the intermode beat signal (Fig.2b), at the fundamental repetition frequency of the comb, is limited to 1.0 Hz by the resolution of the radio-frequency spectrum analyzer. The two pulse trains are amplified and they are launched, for spectral broadening [17], into a nonlinear optical fiber of high normal dispersion.

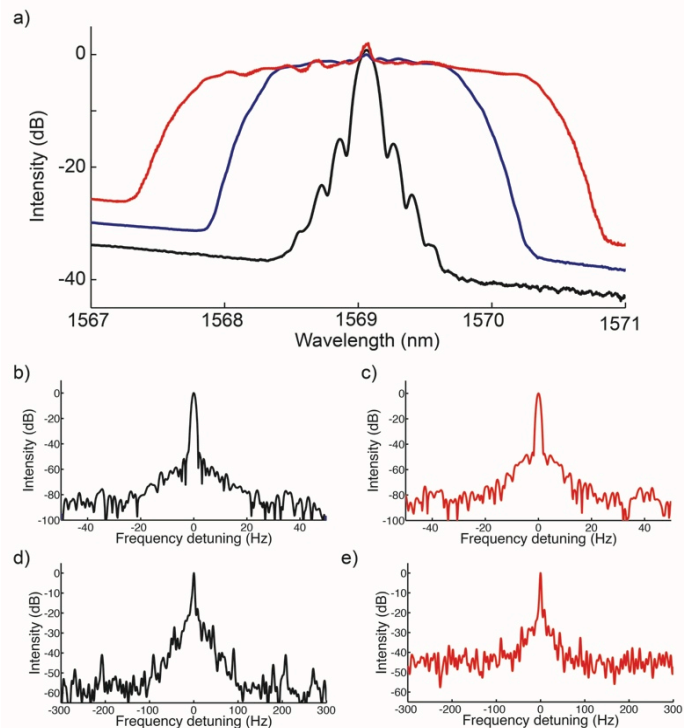

**Figure 2. Characterization of the frequency combs generated by wave-breaking in a fiber of normal dispersion.**
a) Low-resolution optical spectra at the input and output of the normal-dispersion optical fiber. The seed spectrum (black) at the input of the fiber has a FWHM of 0.13 nm. At its output, it spans 2nm (blue) with a launch average power of 17 dBm (50 mW, peak power: 3.33 W). The span broadens to 3 nm (red) for an average power of 24 dBm (250 mW, peak power: 16.75 W).
(b,c) Intermode beat signal at the fundamental repetition frequency (300 MHz) of the comb at the input and at the output of the nonlinear fiber, respectively.
(d,e) Beat note between a pair of optical lines of the two combs, resolved by a fast photodiode and a radio-frequency spectrum analyzer, at the input and the output of the fiber, respectively. The lines are located thirty lines away from the continuous-wave laser carrier.

Generated by wave-breaking [18,19] in the fiber, two flat-top frequency combs, of slightly different line spacing, span (Fig.2a) 3 nm (365 GHz). The generation of combs of large repetition frequency (>10 GHz) using wave-breaking had been reported [20,21] for radio-frequency photonics and arbitrary waveform generation applications, with a set-up involving cascaded electro-optic intensity and phase modulators and a nonlinear fiber of low normal dispersion. The intermode beat signal at the output of the fiber (Fig.2c) is similar to that at the input of the fiber. The comb spectra comprise more than 1350 individual lines, with a power per line of 0.18 mW. Such number of comb lines is nearly 30-fold higher than that generated by a single dual-drive Mach-Zender modulator [12] and this proves essential for simultaneously interrogating several transitions. For efficient common-noise rejection and for enhancing the coherence time between the two combs, we use a single nonlinear fiber, in which the two pulse trains counter-propagate and are separated with a circulator at each end of the fiber. A coupler combines then the two beams. An optical filter may be used to reduce the span. One output of the coupler passes through an absorption cell, whereas the second arm serves as a reference. In the frequency domain, each pair of optical lines, one from each comb, produces a radio-frequency beat note on a single photodetector.

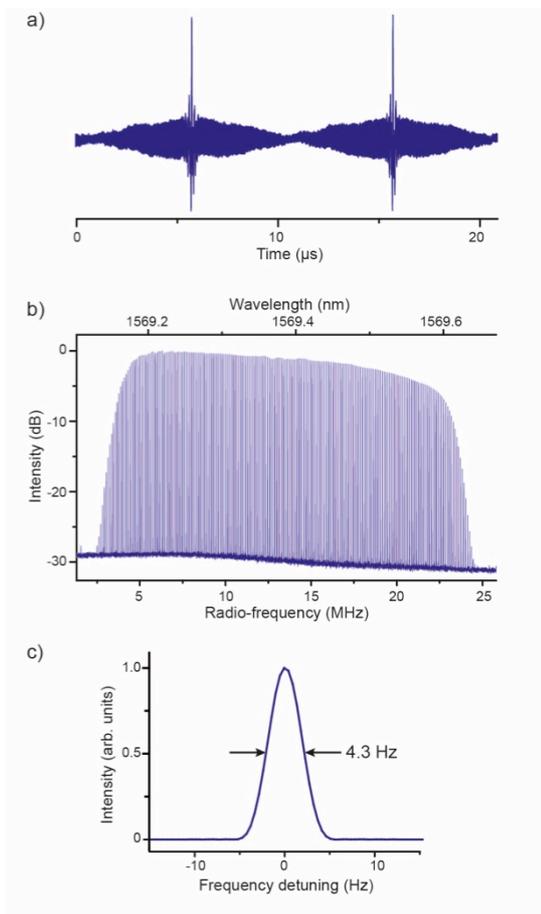

**Figure 3. Experimental interferogram and spectra with resolved comb lines.**
a) Time-domain interferogram. The bursts in the interferogram reproduce with a periodicity of 9.7 µs, the inverse of the difference in comb line spacing (103 kHz). They are due to the low-resolution envelope of the comb emission while the modulation between two such bursts is due to the absorbing molecular lines.
b) Dual-comb spectrum on a logarithmic intensity scale, without absorber.
c,d) Zoom on a resolved comb line after Fourier transformation of an interferogram of 33554432 samples. The instrumental width of the comb line is 4.3 Hz in the radio-frequency domain (13 kHz in the optical domain), although the resolution in the absorption spectrum is limited by the comb line spacing $f$ = 300 MHz.

A frequency comb in the radio frequency region is thus formed by down-conversion of optical frequencies. An individual radio-frequency line resulting from the beat between two optical comb lines, one from each comb, is measured with a radio-frequency spectrum analyzer at the input (Fig.2d) and the output (Fig.2e) of the nonlinear fiber. It exhibits a FWHM of 2.5 Hz, demonstrating a relative coherence time exceeding 400 ms. Despite the kilometric length of the nonlinear fiber, coherence between the two combs is thus satisfactorily maintained, mostly because the same fiber is employed for the spectral broadening of the two combs.

For spectroscopic experiments, the time-domain interference signal at the photodetector is sampled by a fast digitizer and Fourier transformed, allowing for multiplex measurements. Here, the difference in line spacing between the two 300-MHz combs is set to 103 kHz and would make it possible to measure a total optical span exceeding 3 nm without aliasing. Using narrow spectral spans has however two distinguishing features that become manifest in an experimental interferogram (Fig.3a). First, as the ratio between the most intense and the weakest interferometric samples is about one thousand, the dynamic range required to amplify and digitalize the signal is reasonable. Such conditions are favorable to obtaining high signal-to-noise ratio within a short time despite the limited resolution of fast digitizers. The strategy of spectrally filtering a broadband light source to improve the signal-to-noise ratio is well known [20] in Fourier transform spectroscopy and has already been employed [9,10] in dual-comb spectroscopy with mode-locked lasers. The novel approach of harnessing frequency-agile frequency combs of moderate span exploit similar arguments, but with significantly enhanced versatility and simplicity. Second, the interferogram reproduces with a periodicity of 9.7 μs. Such fast refresh times are promising in conjunction with frequency-agile frequency combs, for instance if time-varying processes are studied.

The Fourier transform of an interferogram reveals a radio-frequency spectrum, which can be a posteriori calibrated on the optical scale (Fig.3b). The spectral span of the filter is chosen to 0.5 nm (60 GHz). The spectrum shows 200 resolved lines with a signal-to-noise ratio exceeding 2000. With an interferometric sequence of a total duration of 540 ms, the FWHM of the individual comb lines (Fig.3c) is 4.3 Hz in the radio-frequency spectrum (13 kHz in the optical domain). For comparison, the width of an individual comb line of a free-running mode-locked erbium-doped fiber laser was found [6] to be 260 kHz over an integration time of 1.3 s. Our system of flat-top dual-combs without phase-lock electronics is therefore suitably designed for applications involving a coherent interferometer. Systems involving two lasers [5-11] have been harnessed most of the time in dual-comb spectroscopy, except for a recent demonstration [12]. The demanding relative stability (the beat note between a pair of comb lines, one from each comb, should be narrower than the inverse of the acquisition time) between the two combs is challenging to achieve with independent lasers. Furthermore, with mode-locked lasers, the high peak power of the pulses, the broad spectral bandwidth and the possible large intensity variations in the spectrum induce nonlinearities and limited signal-to-noise ratio.

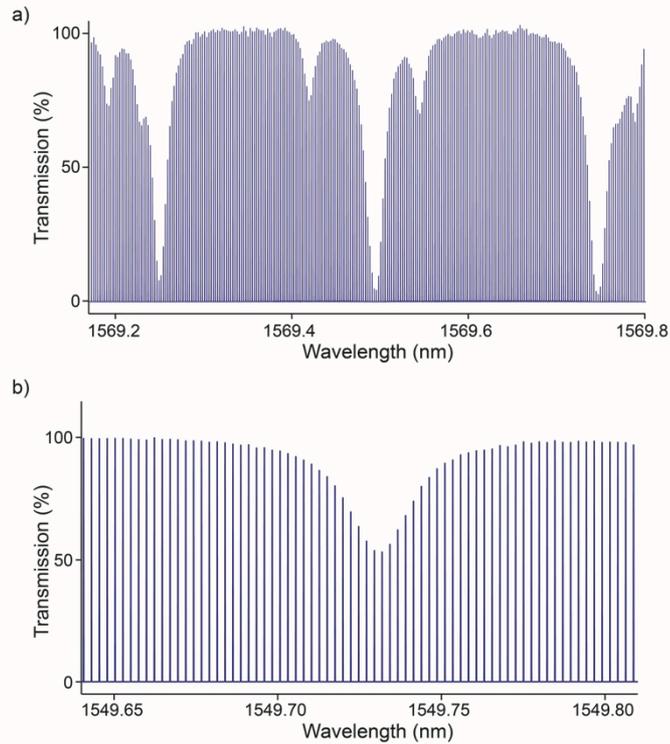

**Figure 4. Portions of experimental dual-comb molecular spectra.**
One hundred spectra, each with a recording time of 524 µs, are averaged. The FWHM of the individual comb lines– 1.9 kHz – is transform limited.
a) Gas mixture containing 90% of $^{12}C^{16}O_2$ and 10% of $^{13}C^{16}O_2$ at a total pressure of 25000 Pa in a hollow-core photonic crystal fiber. The fiber has a length of 48 m, an attenuation less than 30 dB.km$^{-1}$, a mode diameter of 9 µm, and is specified from 1490 to 1680 nm. It is enclosed in a home-built vacuum-tight tank. The ro-vibrational transitions are assigned to $^{12}C^{16}O_2$, 31112-01101 band, R(23) line; $^{12}C^{16}O_2$, 31112-01101 R(22); $^{12}C^{16}O_2$, 30012-00001, R(38); $^{13}C^{16}O_2$, 30011-00001, R(10); $^{12}C^{16}O_2$, 31112-01101, R(21); $^{12}C^{16}O_2$, 30012-00001, R(36); $^{12}C^{16}O_2$, 31112-01101 R(20); $^{12}C^{16}O_2$, 30012-00001, R(34); $^{13}C^{16}O_2$, 30011-00001, R(8) and $^{12}C^{16}O_2$, 31112-01101 R(19).
b) P(10) line of the $2\nu_3$ band of H$^{13}$CN. The fibered gas cell has an absorption path length of 16.5 cm and a gas pressure of 3300 Pa.

We first illustrate the potential of our system for molecular spectroscopy with the weak absorption of the overtones lines of $CO_2$ around 1570 nm. A hollow-core fiber [23] of a length of $L_{abs}$=48 m is filled with a gas mixture of $^{12}C^{16}O_2$ and $^{13}C^{16}O_2$. A portion of a spectrum is shown in Fig.4a. One of the lines has an intensity as small as 6 10$^{-25}$ cm.molecule$^{-1}$. The noise-equivalent-absorption coefficient at 1s-time-averaging per comb line, defined as $(L_{abs}\ SNR)^{-1}(T/M)^{1/2}$, is 1.7 10$^{-9}$ cm$^{-1}$.Hz$^{-1/2}$, where $SNR$ (=2600) is the signal-to-noise ratio, $T$ (=52.4 ms) the measurement time and $M$ (=115) the number of comb lines. The long optical path between the gas and the light guided-mode field in the small volume of the hollow core fiber benefits high sensitivity absorption spectroscopy without multi-pass cell. Tuning the laser diode, without other adjustments, makes it possible to explore a different spectral region. Currently this capability is limited by the employed laser diode to a tuning speed of 10 nm.s$^{-1}$, but faster speeds are commercially available. The spectrum of the P(10) line of the $2\nu_3$ band of H$^{13}$CN in a fiber-coupled cell, around 1550 nm, is measured (Fig. 4b), within 52.4 ms, with a signal-to-noise ratio of 2500. The wavelength-scale calibration is straightforward, as it directly derives from the measurement of the laser diode wavelength and from the comb line spacing, set by radio-frequency generators.

The measurement time of a spectrum may be as short as 12.7 µs, as illustrated (Fig.5a) with the P(9) line of the $2\nu_3$ band of $H^{13}CN$, which is satisfactorily adjusted to a Voigt profile. Spectroscopic monitoring of a couple of Doppler-limited transitions of a dynamically evolving single-event could therefore be performed at an acquisition rate exceeding 80 kHz.

For an entire spectroscopic validation of the experiment, we record spectra of the well-characterized $\nu_1+\nu_3$ band of $C_2H_2$. The experimental profile of its P(23) ro-vibrational line (Fig.5b) at 1539.43 nm is compared to a computed spectrum, which uses the line parameters available in the HITRAN database [24]. The residuals between the experimental and calculated spectra have a standard deviation of 0.6% and do not show any systematic deviation. Such agreement confirms the appropriateness of our technique for the measurement of line profiles and concentrations.

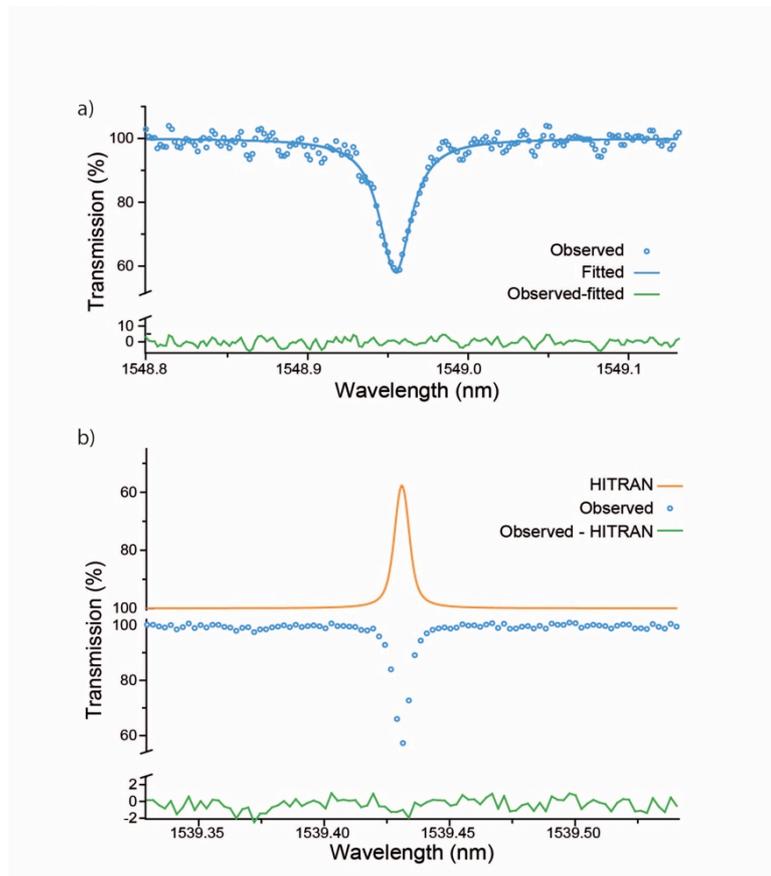

**Figure 5. Short measurement times and spectroscopic validation.**
The ratio of the maxima of the comb lines in the absorption and reference spectra is plotted with dots. Note that the y-scale of transmission only goes down to about 50%.
a) Portion of a spectrum showing the P(9) line of the $2\nu_3$ band of $H^{13}CN$. The measurement time is 12.7 µs. The resolution is 300 MHz. The experimental profile is satisfactorily fitted by a Voigt function. The standard deviation of the residuals is 2.2%, at the noise level.
b) Comparison between the experimental spectrum of the P(23) line of the $\nu_1+\nu_3$ band of $C_2H_2$ and a profile computed using the line parameters available in HITRAN and a Voigt profile. The fibered cell has an absorption path length of 5.5 cm and a gas pressure of 6666 Pa. The temperature is 296 K. One hundred spectra, each measured within 524 µs, are averaged.

We demonstrate a compact dual-comb system, which explores the original association of fast frequency tuning and simultaneous analysis of moderate spectral spans on a single photodetector. No active stabilization or phase-lock electronics is required and, predominantly, passive optical components for the telecommunication industry are harnessed. Well-resolved comb lines render the instrumental line-shape negligible. With further system development, improved relative coherence between the combs, better signal-to-noise ratio and higher frequency agility may be reached. The line spacing of the generated combs only depends on electronic settings and does not encompass resonant optical elements. As for other dual-comb systems, the absence of moving parts, the spatial coherence of the laser beams and the frequency precision are part of the interesting characteristics of our spectrometer. Our capabilities for high-speed sensitive multi-heterodyne spectroscopy can be transferred to other spectral regions by nonlinear frequency conversion. Alternatively, the high power per comb line makes it possible to harness nonlinear phenomena, like coherent Raman effects [7]. Our technique thus opens up new opportunities for real-time spectroscopic diagnostics and maybe even for microscopic hyperspectral imaging.

**Acknowledgements**
We thank J. Fatome, G. Fanjoux and P. Morin for advices during the initial experimental work. Financial support by IXCORE Fondation pour la Recherche, PARI PHOTCOM Région Bourgogne, Labex ACTION, FP7-ERC-Multicomb (Grant 267854) and the Munich Center for Advanced Photonics is acknowledged.